\documentclass{pinchcr}

\usepackage[T1]{fontenc}
\usepackage{tikz}
\usepackage[utf8]{inputenc}
\usepackage{amsmath}
\usepackage{amsbsy}

\usepackage{graphicx}
\usepackage{setspace}
\usepackage{esint}
\renewcommand{\vec}[1]{{\bf #1}}
\makeatletter

\usepackage[pdftex]{hyperref}

\newcommand{\grad}\nabla
\makeatother
\usepackage{wasysym}

\title{Forces in dry active matter}
\author{{\bf Ydan Ben Dor}}
\affiliation{Department of Physics and the Russell Berrie Nanotechnology Institute, Technion -- Israel Institute of Technology, Haifa 32000, Israel}
\author{{\bf \underline{Yariv Kafri}}}
\affiliation{Department of Physics, Technion -- Israel Institute of Technology, Haifa 32000, Israel}
\author{{\bf Julien Tailleur}}
\affiliation{Laboratoire Matiere et Systemes Complexes, UMR 7057 CNRS/P7, Universite Paris Diderot, 10 rue Alice Domon et Leonie Duquet, 75205 Paris cedex 13, France}

\begin{document}

\maketitle
\tableofcontents

\maintext
\chapter{Forces in dry active matter}

Depending on one's background, the word pressure brings about
different definitions. Mechanics, energetics, or continuum mechanics
all lead to equally natural expressions for the pressure. For systems
in thermal equilibrium, the choice of definition is immaterial and
measuring or calculating the pressure using the different expressions
gives identical results. In contrast, active systems are out of
equilibrium and it is \textit{a priori} not obvious that the knowledge
gained from equilibrium systems can be applied to active
matter. Indeed, in recent years it has become clear that much of the
equilibrium intuition cannot be exported to even the simplest classes
of active
systems~\shortcite{Marchetti2013RMP,Cates2015MIPS,Bechinger2016RMP}. In
particular, the forces exerted by active systems exhibit features
which are very different from those of equilibrium
systems~\shortcite{Mallory2014,Takatori2014,Yang2014,Fily2014,Solon2015NatPhys,Solon_interactions,Winkler2015SoftMatter,smallenburg2015swim,yan2015force,yan2015swim,ginot2015nonequilibrium,Nikolai2016PRL,speck2016ideal,joyeux2016pressure,falasco2016mesoscopic,fily2017mechanical,rodenburg2017van,sandford2017pressure,razin2017forces,razin2017generalized,ginot2018sedimentation,sandford2018memory,baek2018generic,Rohwer2018nonequilibrium}.

As we discuss here, this has many implications for systems ranging
from shaken granular gases~\shortcite{junot2017active} to the motion
of cells~\shortcite{Poujade2007}. For example, in equilibrium systems,
in order to change the force exerted by the system on its container
one has to change the {\it bulk} properties of the system. This is a
direct consequence of the existence of an equation of state, which
expresses the pressure solely in terms of bulk quantities of the
system. In contrast, for many active systems the forces exerted by the
system on the container walls do not obey an equation of
state~\shortcite{Solon2015NatPhys}. This implies that forces exerted
by a system on its container can be changed {\it without} changing its
bulk properties but instead by altering the surface of the container.

The purpose of these lecture notes is to give a pedagogical
introduction to recent developments in the thermomechanics of active
systems. Complementary details and more exhaustive treatments can be
found in the corresponding
publications~\shortcite{Solon2015NatPhys,Nikolai2016PRL,fily2017mechanical,baek2018generic}. The
lectures start in Section~\ref{sec:definitions} with a short overview
of different definitions of pressure and the conditions under which
they are equivalent. This allows us to identify a class of systems --
commonly referred to as dry active matter -- where the simple
equilibrium intuition might break. Following this, we turn to discuss
simple models of dry active systems in Section~\ref{sec:overdamped}
and show that there is, in general, no equation of state for the
forces they exert on their container. To better understand the
physical origin of the lack of an equation of state, we then focus in
Section~\ref{sec:steady-state} on an exactly solvable one-dimensional
model. We rationalize our results in Section~\ref{sec:impulse} in terms
of an active impulse, which helps us explain why in certain classes of
finely tuned models an equation of state is restored. Finally, we go
beyond the question of the equation of state and provide a brief
discussion of recent developments concerning the forces exerted on
objects immersed in  active systems in Section~\ref{sec:currents}.

\section{A short recap of different expressions for pressure} \label{sec:definitions}
\begin{figure}
  \begin{center}
    \includegraphics[width=.5\textwidth]{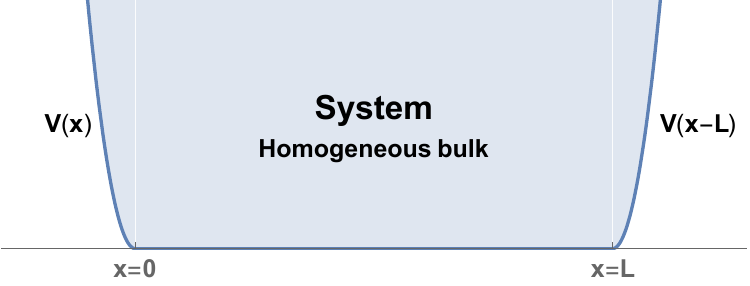}
  \end{center}
  \caption{Illustration of a one-dimensional system of length $L$, confined by two potentials at its two ends. The potentials model the presence of a confining vessel.}\label{fig:confining potential}
\end{figure}

We now present three standard definitions of pressure and show that
they are equivalent in equilibrium. For simplicity, the discussion
will be carried out in one dimension for a system confined between two
walls, specified by bounding potentials at its two ends (see
Fig.~\ref{fig:confining potential}). Following this, we will argue why
things might be different for active systems.

\subsection{An energetic definition} 

In thermodynamics, the pressure is directly related to the change in the energy of a system under a change in its length, $L$, through
\begin{equation} \label{eq:energetic P1}
	P_E=-\frac{\partial E}{\partial L}\Bigr\vert_{S,N}
\end{equation}
with $S$ the entropy of the system (implying here no exchange of heat
with the environment), and $N$ the number of particles in the
system. Of course, other choices of state functions are possible. For
example, using the Helmholtz free energy, $F$, gives
$P_E=-\frac{\partial F}{\partial L}\Bigr\vert_{T,N}$ with $T$ the
temperature at which the system is held fixed as its length
varies. Note that since the energy is extensive, i.e. $E$ is
proportional to $L$, this definition directly implies that the details
of the interactions of the particles with the confining walls, which
in one dimension scales as $L^0$, do not change the pressure. This
insensitivity to the details of the wall potential implies that the
pressure is a state-function that depends only on intensive bulk
quantities. Whether the interaction of the particles with the wall is
hard-core or the interaction binds the particles strongly to the wall, the pressure remains
unchanged.

\subsection{A mechanical definition}
\label{sec:mecdef}
 The second definition of pressure is simply the force per unit area on the walls confining the system. Given a steady-state density of particles at position $x$, $\rho(x)= \sum_{i=1}^{N} \delta(x-x_i)$, with $x_i$ the position of particle $i=1 \ldots N$, the pressure can be written as 
\begin{equation} \label{eq:mechanical_pressure}
	P_M=\int_{x^*}^\infty dx\,\langle \rho(x) \rangle \partial_x V(x-L)\ ,
\end{equation}
where $V(x-L)$ is the potential of the confining wall which, say, starts at position $L$ (see Fig. \ref{fig:confining potential}), the angular brackets denote an average over the steady-state distribution, and $x^*$ is a point deep in the system. By ``deep'' we mean that the steady-state distribution at $x^*$ is independent of the details of the bounding potential\footnote{In practice, this means that $L-x^*\gg \xi$ and $x^* \gg \xi$, where $\xi$ is the correlation length, which is assumed to be finite. }.

In contrast to the energetic definition, the existence of an equation of state (EOS) for the pressure $P_M$, namely the independence on the shape of the confining wall, is not at all obvious -- there is no reason for the integral in~\eqref{eq:mechanical_pressure} to be independent of $V$ \emph{a priori}. However, for equilibrium systems, this is easy to verify by proving the equality $P_M=P_E$. The thermodynamic definition, $P_E$, is given by
\begin{equation}\label{eq:energetic P2}
    P_E=-\frac{\partial F}{\partial L}\Bigr\vert_{T,N} \;,
\end{equation}
with
\begin{equation}
F=-\frac{1}{\beta} \ln {\mathcal Z} \;,
\end{equation}
and
\begin{equation}
{\mathcal Z} = \sum_{\mathcal C} e^{-\beta [ {\mathcal H} + \sum_i V(x_i-L)]} \;.
\end{equation}
Here the sum is over micro-states, $\beta$ is the inverse temperature, and ${\mathcal H}$ contains all the other interactions in the system. Using the definition of $P_E$, as given in Eq. (\ref{eq:energetic P2}), we have
\begin{equation}
P_E=-\frac{1}{{\mathcal Z}} \sum_{\mathcal C} \sum_j \partial_LV(x_j-L) e^{-\beta ( {\mathcal H} + \sum_i V(x_i-L))} =  \int dx \langle \rho(x) \rangle \partial_x V(x-L)=P_M 
\end{equation}
where the angular brackets denote a thermal average. Note that the proof only requires that the system is in equilibrium through the Boltzmann distribution. As stated above, however, there is no \textit{a priori} obvious reason why $P_M$ should in general be independent of $V$.

\subsection{A momentum-flux-based definition} 
\label{sec:mfbdef}

A third definition arises in continuum mechanics. We consider a
momentum-conserving system comprising $N$ particles of mass $m$,
positions $x_i$, and velocities $v_i$. Assuming a local dynamics, the
conservation equation for the momentum field, $\hat{p}=\sum_i
mv_i\delta(x-x_i)$, reads
\begin{equation}\label{eq:boring}
\partial_t {\hat{p}}(x) = -\partial_x J_{\hat{p}}(x)\;.
\end{equation}
Here $J_{\hat{p}}(x)$ is the momentum current which is in general a
noisy fluctuating quantity. For a non-interacting ideal gas, for
example, the momentum flux is given by $J_{\hat{p}}=\sum_i m {v}_i
{v}_i \delta(x-x_i)$, with $v_i$ the velocity of particle $i$ and $m$
the particle mass. One then identifies the pressure with the
steady-state flux of momentum in the bulk of the system when the fluid
is static
\begin{equation}\label{eq:Pp}
	P_p = \langle J_{\hat{p}}(x^*) \rangle = -\sigma \;,
\end{equation} 
where we introduced the stress $\sigma$ and $x^*$ is again a point
deep in the bulk of the system. In higher dimensions,
Eq.~\eqref{eq:Pp} is generalized to the trace of the stress tensor
divided by the number of dimensions,
$P_p=-\frac{1}{d}\mathrm{Tr}\sigma$. The equivalence with the
mechanical definition is rather intuitive for momentum conserving
systems. It can be seen by explicitly including a wall potential in
the equations for the momentum flux
\begin{equation}\label{eq:conservation}
\partial_t {\hat{p}}(x) = -\partial_x {J}_{\hat{p}}(x) - \rho(x)
\partial_x V(x)\;,
\end{equation} 
with $\rho(x)$ the density of particles at point $x$ as defined above. In the steady state this becomes
\begin{equation}\label{eq:froce_balance_conser}
	 -\langle \partial_x {J}_{\hat{p}}(x) \rangle = \langle \rho(x) \partial_x V(x) \rangle\ ,
\end{equation}
which states that the change in momentum flux is caused by the force exerted on the particles by the potential $V$.
Integrating over space we obtain
\begin{equation}
P_p = \langle J_{\hat{p}}(x^*) \rangle =\int_{x^*}^\infty dx \, \langle \rho \rangle\, \partial_x V(x) = P_M \;.
\label{eq:pressure_definition}
\end{equation}
Note that this result implies that any momentum conserving system with
local dynamics (whether in equilibrium or not) and whose bulk
properties are  independent of the bounding potential admits an
equation of state\footnote{Clearly, a trivial non-homogeneity such as
  phase separation is allowed.}. Namely, the force exerted on the wall
(or the momentum flux as defined above) is independent of the shape of
the potential.

\subsection{Discussion}
To summarize the above discussion, we make the following points and observations:
\begin{itemize}
\item The independence of the mechanical pressure on the shape of the potential is guaranteed for both equilibrium systems (Sec. \ref{sec:mecdef}) and systems which conserve momentum and have local dynamics (Sec. \ref{sec:mfbdef}).
\item Depending on the system, either one of these might be at
  play. For example, osmotic pressure allows for an extra
  mechanism of loss of momentum by a flow of the solvent
  through the membrane which is not captured by
  Eq. (\ref{eq:conservation}). In this case, $P_M$ is
  guaranteed to be given by an equation of state only when the
  system is in equilibrium. Conversely, in a non-equilibrium
  system which conserves momentum in the bulk, the mechanical
  pressure exerted on a container will obey an equation of
  state.
\item In active systems which continuously absorb and
  dissipate energy, the thermodynamical definition of pressure
  is clearly useless.
\end{itemize}

In active matter, many experimentally relevant situations involve
systems which do not conserve momentum and are out of
equilibrium. These can be, for example, active particles moving in 2D
next to a surface. Such systems are typically referred to as dry
active systems (as opposed to wet
ones)~\shortcite{Marchetti2013RMP}. The above discussion suggests that
in these systems the existence of an equation of state is not
obvious. Note that the same holds for wet systems confined by porous
walls. Whether or not (and when) the intuition built-in equilibrium on
the statistical properties of forces extend to such active cases is
one of the issues that these lectures address. While the discussion
will be carried out for a particularly simple class of active systems,
the broad results and conclusions, as should be evident, are expected
to hold generally. For pedagogical purposes, we focus on deriving
results for non-interacting particles. Generalizations to cases with
interactions will be commented on and can be found in the
literature~\shortcite{Takatori2014,Yang2014,Solon2015NatPhys,Solon_interactions,falasco2016mesoscopic,rodenburg2017van,fily2017mechanical}.

\section{The mechanical pressure of non-interacting self-propelled particles} \label{sec:overdamped}
\subsection{Active Brownian Particles -- a simple model for active systems}

In what follows we study a model of non-interacting Active Brownian
Particles (ABPs) (all our results directly extend to run-and-tumble
particles (RTPs)). We first introduce and discuss the active dynamics
in the bulk, and then turn to the effect of the walls.

ABPs are described using their positions, $\vec{r}_i$, and
orientations, $\theta_i$, where $i$ is the particle index. In the
overdamped regime, their dynamics are given by:
\begin{eqnarray} \label{eq:motion_no_wall}
    \dot{\vec{r}}_i &=& v\vec{u}(\theta_i) + \sqrt{2D_t}\boldsymbol{\eta}_i(t)\nonumber\ , \\
    \dot{\theta}_i &=& \sqrt{2D_r}\xi_i\;.
\end{eqnarray}
Here $v$ quantifies the activity, signalling that an energy source is constantly used in order to propel the particle. $\vec{u}(\theta_i) = (\cos(\theta_i),\sin(\theta_i))$ is a director, $D_t$ and $D_r$ are the translational and rotational diffusion coefficients, respectively, $\boldsymbol{\eta}_i,\, \xi_i$ are random Gaussian noises of zero mean and correlations $\langle \eta_i^k(t) \eta_j^\ell(t') \rangle = \delta_{ij}\delta^{k\ell}\delta(t-t')$, where $k$ and $\ell$ refer to spatial components, and $\langle \xi_i(t) \xi_j(t') \rangle = \delta_{ij} \delta(t-t')$. It is obvious that this system does not conserve momentum and is therefore a dry active system. 

\begin{figure}
  \begin{center}
    \includegraphics[width=.4\textwidth]{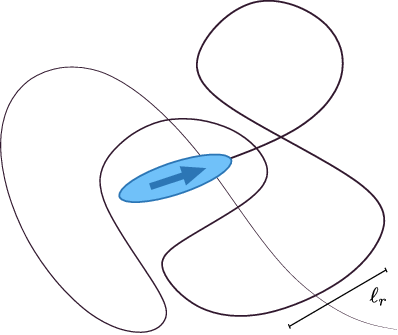}
  \end{center}
  \caption{A cartoon trajectory of an active Brownian particle. The
    particle is self-propelled at constant speed $v$; it undergoes
    translational and rotational diffusion. The particle reorients
    itself after a persistence length of $\ell_r=v/D_r$.}\label{fig:ABP
    trajectory}
\end{figure}

The outcome of these dynamics is that an active Brownian particle roughly follows a straight trajectory over a persistence length $\ell_r = \frac{v}{D_r}$ before reorienting itself (as depicted in Fig. \ref{fig:ABP trajectory}). Therefore, on large length scales and long times, an ABP has a diffusive dynamics. The diffusion constant can be obtained through a simple dimensional analysis. It is given by
\begin{equation}
    D_{\rm eff} = \frac{v^2}{d D_r} = \frac{1}{d} v \ell_r
\end{equation}
with $d$ being the spatial dimension of the system.

Note that since each particle is self-propelled in a given direction,
looking at a movie showing an ABP moving around is different from
looking at its reversed version: running the movie backward will not
change the director's direction. This means that ABPs break
time-reversal symmetry in a trivial manner, which shows that this
model is out of equilibrium\footnote{Note that the breakdown of
  time-reversal symmetry is much less apparent if only the position of
  the particle is recorded, and not its
  orientation~\shortcite{Fodor2016PRL,Mandal2017PRL,Shankar2018PRE}.}.

We now complement this basic model to include the presence of
confining walls and calculate the pressure $P_M$.

\subsection{The pressure of ABPs}\label{sec:Pressure of ABPs}

\begin{figure}
  \begin{center}
    \includegraphics[width=.5\textwidth]{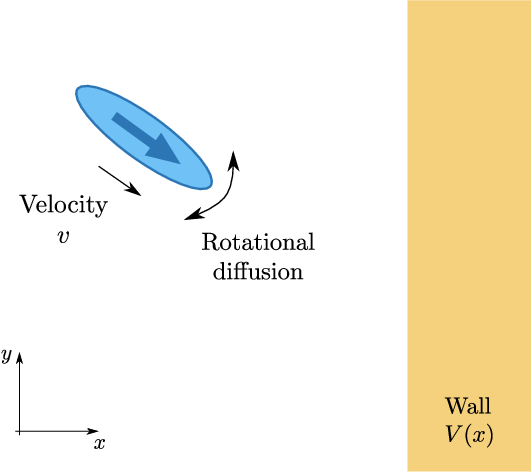}
  \end{center}
  \caption{An active Brownian particle next to a wall. The particle is
    confined in the $\hat x$ direction whereas periodic boundary
    condition are used along $\hat y$. The wall is modelled by a
    potential $V(x)$ that leads to a confining force, and may also lead to
    torques acting on the particle.}\label{fig:ABP near a wall}
\end{figure}

To calculate $P_M$, we consider a system of $N$ particles, $i=1,...,N$
that follow the equations of motion \shortcite{Solon2015NatPhys}
\begin{eqnarray} \label{eq:motion_withwall}
    \dot{\vec{r}}_i &=& v\vec{u}(\theta_i) - \mu \nabla V + \sqrt{2D_t}\vec{\eta}_i(t) \nonumber\ ,\\
    \dot{\theta}_i &=& \Gamma (\vec{r}_i,\theta_i) + \sqrt{2D_r}\xi_i \;.
    \label{eq:ABP wall model}
\end{eqnarray}
Here, on top of the dynamics of Eq. (\ref{eq:motion_no_wall}), we introduce a confining wall modelled by a potential $V$. In what follows, we take the potential to be uniform along the $\hat{y}$ direction (Fig. \ref{fig:ABP near a wall}), with periodic boundary conditions. The dynamics of the angles $\theta_i$ now accounts for torques acting on the particles due to the wall through the term $\Gamma(\vec{r}_i,\theta_i)$. Evidently, $\Gamma$ vanishes in the bulk and, with the exception of circular particles, is non-zero in the vicinity of the wall\footnote{Note that for ease of notation the rotational mobility, $\mu_r$, is absorbed into $\Gamma$.}. Finally, the noise terms are taken identical to the case without a wall -- white noises with unit variance. We stress again that this model does not conserve momentum and is not in thermal equilibrium (for $v\neq 0$). This is the class of models where interesting behavior might be expected.

To calculate $P_M$, we first write the Fokker-Planck equation corresponding to Eq. (\ref{eq:motion_withwall})
\begin{eqnarray}
    \partial_t \mathcal{P}(\vec{r},\theta) &=& -\nabla \cdot [v\vec{u}(\theta)\mathcal{P}(\vec{r},\theta) - \mu (\nabla V) \mathcal{P}(\vec{r},\theta) - D_t\nabla \mathcal{P}(\vec{r},\theta)] \nonumber \\ &&- \partial_\theta [\Gamma (\vec{r},\theta) \mathcal{P}(\vec{r},\theta) - D_r \partial_\theta \mathcal{P}(\vec{r},\theta)] \;.
\label{eq:FP ABP}\end{eqnarray}
Since the particles are non-interacting, $\mathcal{P}(\vec{r},\theta)$ can be identified with the average density of particles at position $\vec{r}$ with angle $\theta$. Integrating Eq. (\ref{eq:FP ABP}) over $\theta$ and utilizing the translational symmetry along the $\hat{y}$ direction, we obtain
\begin{equation}
      \partial_t\rho(\vec{r}) = -\partial_x J_x(\vec{r}) \;\;\; ; \;\;\;
      J_x(\vec{r}) = v m_1(\vec{r}) - \mu (\partial_x V) \rho(\vec{r}) - D_t \partial_x \rho(\vec{r}) \;.
\end{equation}
Here $J_x(\vec{r})$ is the density current along the $\hat x$ direction, 
\begin{equation}
    m_n(\vec{r}) = \int d\theta\, \cos(n\theta)\mathcal{P}(\vec{r},\theta) \;,
\end{equation}
and we denote $\rho(\vec{r})=m_0(\vec{r})$. In the steady state, the geometry of the system implies that $J_x(\vec{r}) = 0$, which sets
\begin{equation}
    (\partial_x V)\rho(\vec{r}) = \frac{1}{\mu}[v m_1(\vec{r}) - D_t \partial_x \rho(\vec{r})]\ .
\end{equation}
Integrating this expression from a point deep in the bulk of the system\footnote{Here we assume that another wall is positioned far away at the left end of the system and ensures the existence of a steady state.} (which we set to be $x=0$) to infinity gives
\begin{eqnarray}
      P_M & = & \int_{0}^{\infty}dx\,\rho(\partial_x V) =  \frac{1}{\mu}\bigg[D_t \rho_0 + \int_{0}^{\infty}dx\, v m_1(\vec{r})\bigg]
    \label{eq:P_M for ABP}
\end{eqnarray}
where $\rho_{0}$ is the density of particles in the bulk of the system. To proceed, we solve for $m_1$. Multiplying Eq. (\ref{eq:FP ABP}) by $\cos(\theta)$ and integrating over the  angle $\theta$, we find that in the steady state
\begin{eqnarray}
      D_r m_1(\vec{r}) & = & -\partial_x\bigg[v \frac{\rho(\vec{r}) + m_2(\vec{r})}{2} - \mu (\partial_x V) m_1(\vec{r}) - D_t \partial_x m_1(\vec{r})\bigg]\nonumber \\
      && - \int_{0}^{2\pi}d\theta\,\Gamma(x,\theta)\sin(\theta)\mathcal{P}(\vec{r},\theta)\ ,
    \label{eq:m1}
\end{eqnarray}
where we used $\cos^2(\theta) = \frac{1}{2}[1 + \cos(2\theta)]$. Using
this in Eq. (\ref{eq:P_M for ABP}) and noting that the system is
isotropic in the bulk so that $\left.m_n\right\vert_{x=0} = 0$, gives
\shortcite{Solon2015NatPhys}
\begin{equation}
    \boxed{P_M = \rho_0\bigg[\frac{v^2}{2\mu D_r} + \frac{D_t}{\mu}\bigg] - \frac{v}{\mu D_r}\int_{0}^{\infty}dx\int_{0}^{2\pi}d\theta\,\Gamma(x,\theta)\sin(\theta)\mathcal{P}(x,\theta)}
    \label{eq:P_M}
\end{equation}

The above result has several implications which we now discuss in detail:

\medskip\noindent\textbf{Equilibrium limit.} In equilibrium, taking $v = 0$, the mechanical pressure is, as expected, given by the ideal-gas result
\begin{equation}
    P_M = \rho_0 \frac{D_t}{\mu} = \rho_0 T
\end{equation}
where we set the Boltzmann constant to unity and use the fluctuation-dissipation relation $D_t=\mu T$.

\medskip\noindent\textbf{Torque-free particles.} In cases where the walls do not exert any torque on the active particles, $\Gamma=0$, the mechanical pressure $P_M$ is independent of the wall potential $V(x)$. In this case, the pressure is given by
\begin{equation}
    P_M =  \rho_0 \left[ T + \frac{1}{\mu}\frac{v^2}{2D_r}\right] \;.
\end{equation}
The first term is the ideal gas contribution, while the second is given by $\rho_0 D_{\rm eff}/\mu$. This second term is what one might guess from dimensional analysis; historically, it was first proposed using continuum mechanics arguments and a virial formula~\shortcite{Mallory2014,Yang2014,Takatori2014,falasco2016mesoscopic}.
    
\medskip\noindent\textbf{The impact of torques and the lack of an
  EOS.}  In the presence of torques, $\Gamma(\vec{r},\theta)\neq 0$
depends explicitly on the functional form of the wall
potential. Furthermore, $\mathcal{P}(\vec{r},\theta)$ depends on it
implicitly as well. This, in general, renders the integral part of the
pressure wall-dependent (see Eq. (\ref{eq:P_M})). Therefore, for such
systems the pressure $P_M$ \textit{does not admit an equation of
  state} -- in order to compute $P_M$, one must specify $V(x)$ and not
solely consider the bulk properties of the active system. We now
illustrate this for several examples and discuss the consequences.

\medskip\noindent\textbf{Point-like elliptical particles.} 
Consider elliptical particles of axes of length $a$ and $b$, with $a$ the length along the propulsion axis, confined by harmonic walls $V(x-L) = \frac{\lambda}{2}(x-L)^2$. In the limit of small ellipses, the torque is given by~\shortcite{Solon2015NatPhys}
\begin{equation}
    \Gamma = \lambda\kappa\sin(2\theta)
\end{equation}
with $\kappa=\frac{1}{8}(a^2-b^2)$. Neglecting translational diffusion ($D_t=0$), one can show that to an excellent approximation (see \shortciteNP{Solon2015NatPhys} for details) the pressure is given by
\begin{equation}
    P_M = \frac{\rho_0 v^2}{2\mu\lambda\kappa}\bigg[1-\exp{\bigg(-\frac{\lambda\kappa}{D_r}\bigg)}
    \bigg]\ .\label{eq:ellipses P_M}
\end{equation}
This dependence is shown in Fig. \ref{fig:P_M for
  ellipses}. Importantly, for both $\kappa$ positive and negative,
there is an order-one change of the magnitude of the pressure as the
strength of the harmonic potential $\lambda$ changes, highlighting the
important impact of wall torques on the pressure.

\begin{figure}
  \begin{center}
    \includegraphics[width=.6\textwidth]{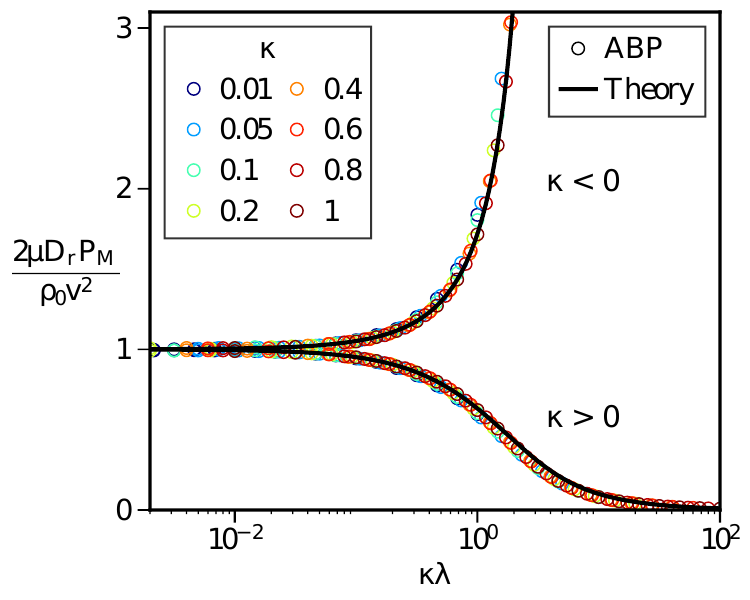}
  \end{center}
  \caption{Pressure exerted by non-interacting, self-propelled
    Brownian ellipses near a harmonic wall as a function of their
    asymmetry. (Figure courtesy of
    Yaouen Fily.) The pressure is normalized by the pressure of
    torque-free particles. The theory corresponds to
    Eq.~(\ref{eq:ellipses P_M}). The two cases plotted in the figure
    correspond to positive and negative $\kappa$. The strong
    dependence of the pressure on $\kappa$ is qualitatively
    explained in Fig.~\ref{fig:wall torques}.}\label{fig:P_M for ellipses}
    \end{figure}
    
\begin{figure}
  \begin{center}
  \includegraphics[width=.8\textwidth]{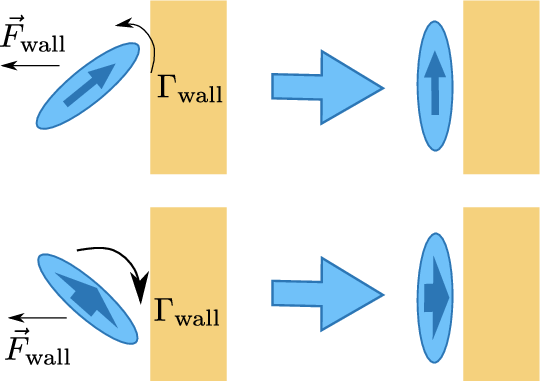}
  \end{center}
  \caption{Impact of wall torques on the active contribution to the
    mechanical pressure. {\bf Top:} For $\kappa>0$, a torque aligns the particle's
    orientation with the \textit{tangent} to the wall, hence diminishing
    its contribution to the pressure. {\bf Bottom:} For $\kappa<0$, a
    torque aligns the particle's orientation with the
    \textit{normal} to the wall, hence increasing the mechanical
    pressure.}\label{fig:wall torques}
\end{figure}

An intuitive illustration of how torques affect the value of $P_M$ is given in Fig. \ref{fig:wall torques}. The active contribution to the pressure arises from the particles transferring their active forces onto the wall. If torques rotate the particles and constrain their motion along the wall, the force acting on the wall is reduced (compared to, say, torque-free particles). Conversely, if torques make particles face the wall, the forces acting on the wall are enhanced.

\medskip\noindent\textbf{Pairwise forces.} Following a more elaborate derivation for a torque-free model of ABPs interacting via pairwise interactions of the form $U_{ij}=U(\vec{r}_i-\vec{r}_j)$, one finds that an EOS exists in this case. The pressure $P_M$ can then be expressed in terms of bulk correlators (see~\shortciteNP{Solon2015NatPhys,Solon_interactions,fily2017mechanical} for details) which coincides with the pressure derived using a virial/continuum mechanics approach~\shortcite{Takatori2014,Yang2014,falasco2016mesoscopic}. 

\medskip\noindent\textbf{Aligning and quorum-sensing interactions.} For models ignoring the torques exerted by the walls but including either aligning interactions or quorum-sensing interactions (in which the single-particle velocity varies with the local density, $\rho(\vec{r})$, in its vicinity, $v=v[\rho(\vec{r})]$), one finds that there is no EOS. Here again, $P_M$ depends explicitly on the form of $V(\vec{r})$.

\begin{figure}
  \begin{center}
    \includegraphics[width=.8\textwidth]{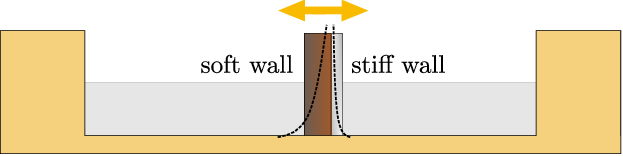}
  \end{center}
  \caption{A mobile partition that is stiffer on one side divides the system into two compartments, each initially with equal density. Since the pressure depends on the wall's stiffness through Eq.~(\ref{eq:ellipses P_M}), each side of the partition will experience a different force from the active particles. As a result, the mobile partition moves until the forces on both sides balance.}\label{fig:piston cartoon}
\end{figure}

\begin{figure}
  \begin{center}
    \includegraphics[width=.49\textwidth]{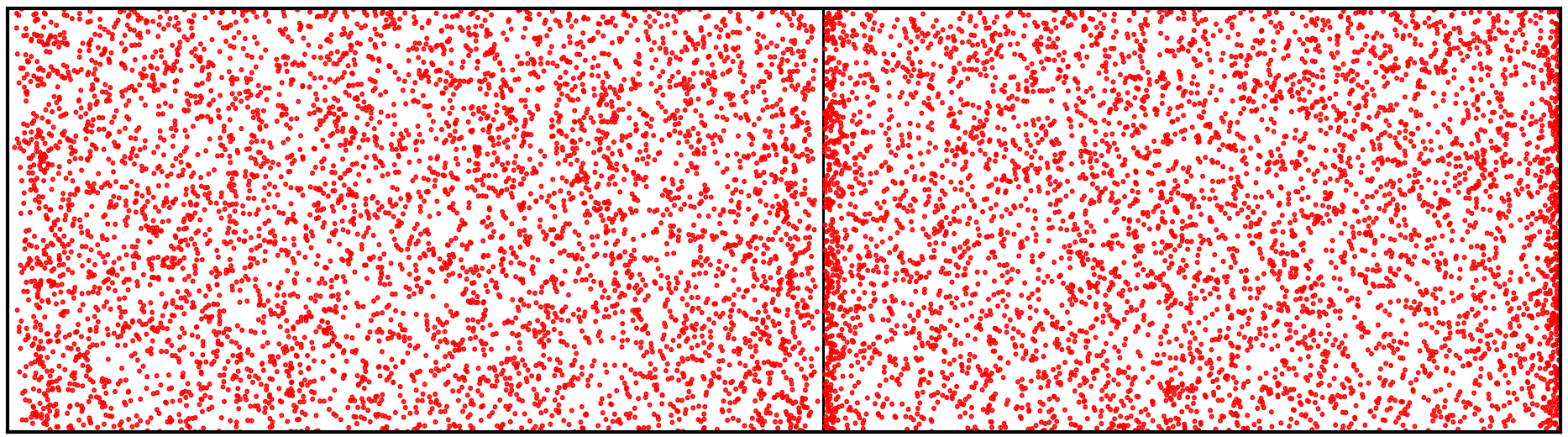}
    \includegraphics[width=.49\textwidth]{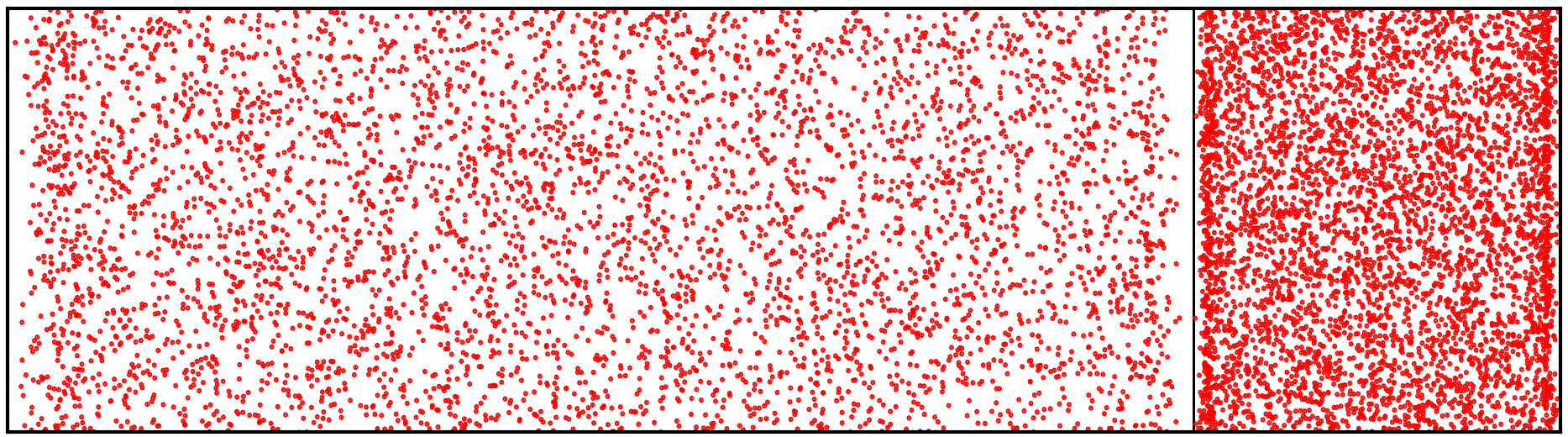}
  \end{center}
  \caption{Numerical simulations corresponding to the setup described
    in Fig~\ref{fig:piston cartoon} using either torque-free ABPs
    (left panel) or elliptical ABPs (right panel). The absence of an
    EOS in the latter case is apparent from the spontaneous
    compression of one half of the system. Figure adapted
    from~\protect\shortcite{Solon2015NatPhys}.}\label{fig:piston}
\end{figure}

\begin{figure}
  \begin{center}
    \includegraphics[width=.5\textwidth]{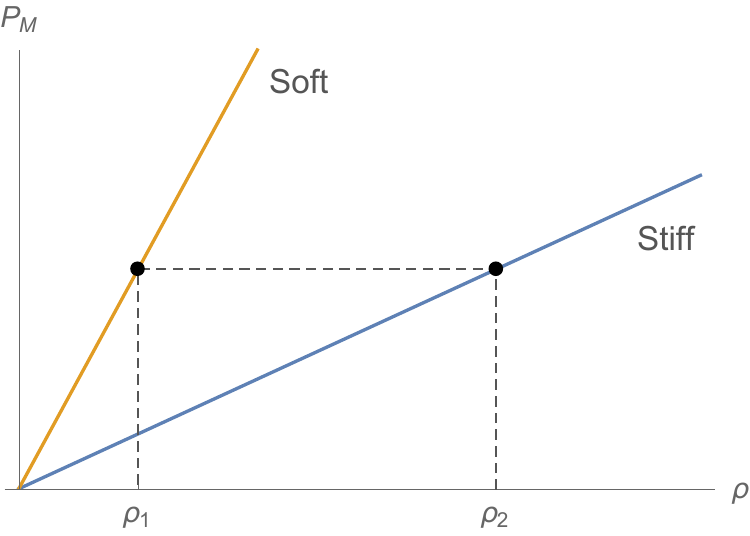}
  \end{center}
  \caption{Pressure vs density for both compartments in Fig.~\ref{fig:piston cartoon}. As evident from Eq. (\ref{eq:P_M}), the pressure is a linear function of the density $\rho$ and a monotonically decreasing function of the stiffness $\lambda$. Therefore, balancing the forces on both sides of the partition, as illustrated in Fig. \ref{fig:piston cartoon}, amounts to finding the piston location, $x$, that satisfies $x\rho_1 + (1-x)\rho_2 = \rho_{\text{initial}}$ with $\rho_1$ and $\rho_2$ the densities of equal pressure, assuming a homogeneous initial state. This explains the behavior shown in Fig.~\ref{fig:piston}.}\label{fig:pressure difference piston}
\end{figure}

\medskip\noindent\textbf{Obstacles with asymmetric stiffness.}
In the large size limit, changing the wall potential has no effect on the {\it bulk} properties of the system. In the presence of an equation of state, $P_M$ is then independent of the wall potential.
On the contrary, when $P_M$ does not obey an equation of state, changing the potential of the walls alters the forces acting on it. This result has recently been observed experimentally in systems of vibrated, self-propelled grains~\shortcite{junot2017active}. This effect can have striking consequences. 
Consider a system of homogeneous active particles in the middle of which a mobile partition is inserted (see Fig.~\ref{fig:piston cartoon}). One side of the partition has a stiffer potential than the other side. The lack of an EOS implies that the partition will move until the densities on both sides are such that the values of the mechanical pressure $P_M$, corresponding to different potentials on each side, are equal (see Fig.~\ref{fig:piston}). The new position of the partition can be obtained using the construction illustrated in Fig. \ref{fig:pressure difference piston} and, in general, leads to unequal densities on both sides of the partition. In fact, this setup is a natural test for the existence of an EOS.

\hfill \break

In the discussion above, we found that in dry active systems, that are out of equilibrium by construction and do not conserve momentum, the pressure $P_M$ may or may not have an EOS. It is interesting to understand what are the conditions needed to ensure the existence of an EOS for $P_M$ and conversely when do these conditions break. This will be the focus of much of the discussion that follows. We will show that when an equation of state exists, it results from a new conservation law which holds statistically in the steady state.

It is obvious from~\eqref{eq:P_M} that the mechanical pressure depends crucially on the form of the steady-state density in the presence of a potential. To study whether the simple fact that this steady state is not given by a Boltzmann law suffices to explain the lack of EOS, we first look at a simple one-dimensional active system, whose steady-state distribution can be computed exactly. Much of the conclusions from this simple example carry over to more complicated settings.

\section{Run-and-Tumble particles in 1D: non-local steady state and equation of state}\label{sec:steady-state}

Consider a one-dimensional system of non-interacting run-and-tumble
particles, originally inspired by the motion of {\it E. Coli} bacteria
\protect\shortcite{berg1993random,schnitzer1993theory,Tailleur2008PRL}. In
the absence of an external potential, particles move with velocity $v$
or $-v$, changing between the two with rate $\alpha/2$. Accounting for
the action of an external potential $V(x)$, the equations for the
average density of right moving particles, $P_+(x,t)$, and left-moving
particles, $P_-(x,t)$, are given by
\begin{eqnarray}
	\partial_t P_+(x,t)&=&-\partial_x \left[ vP_+(x,t)-\mu (\partial_x V) P_+(x,t) \right] -\frac{\alpha}{2}P_+(x,t) +\frac{\alpha}{2}P_-(x,t) \nonumber \ ,\\
	\partial_t P_-(x,t)&=&-\partial_x \left[ -vP_-(x,t)-\mu (\partial_x V) P_-(x,t) \right] -\frac{\alpha}{2}P_-(x,t) +\frac{\alpha}{2}P_+(x,t)\;.
\end{eqnarray}  
Here $\mu$ is the mobility and the translational diffusion of the particles is ignored ($D_t=0$). In the following we follow the derivation presented in~\shortciteNP{Solon2015NatPhys} (See also~\shortciteNP{Hanggi1984} for an  earlier derivation).
While this implies that the particles cannot cross a potential which exerts a force larger than $v/\mu$, this feature does not change the main points that we report below. 

Assuming that the system is confined at its boundaries so that there is no current flowing through the system, the equation for the steady-state density $\rho(x)=P_+(x) + P_-(x)$ reads
\begin{equation}
    \partial_x \left[ (v^2-\mu^2(\partial_x V)^2 )\rho \right]+\alpha \mu (\partial_x V) \rho=0 \;,
    \label{eq:1druntumble}
\end{equation}
whose solution is given by
\begin{equation}
    \rho(x)=\rho(0)e^{-Q}
\end{equation}
with $\rho(0)$ the density at $x=0$, and
\begin{equation}\label{eq:Q}
    Q=\ln\bigg(1-\left(\frac{\mu}{v}\right)^{2}\left(\partial_{x}V\right)^{2}\bigg)
    +\frac{\alpha\mu}{v^{2}}\int_0^x dx' \,\frac{\partial_{x'}V(x')}{\left(\frac{\mu}{v}\right)^{2}\left(\partial_{x'}V\right)^{2}-1} \;.
\end{equation}
This result was first derived in~\shortciteNP{Hanggi1984} in the context of
stochastic flows driven by non-Gaussian noises.

It is also interesting to look at the polarization of the active particles $\Delta(x)=P_+(x)-P_-(x)$. This can be checked to obey the differential equation
\begin{equation}
	\Delta=\frac{1}{\alpha} \partial_x \left(v \rho - \mu (\partial_x V) \Delta \right)
	\label{eq:Delta eom}
\end{equation}
whose consequences we discuss below. 

\hfill \break
The above results imply the following:
\begin{itemize}
    \item To leading order in $V$, $Q(x)=\frac{\alpha\mu}{v^2}V(x)$ ---  the density is then a \textit{local} function  of the potential $V(x)$ and behaves as an effective Boltzmann distribution, with an effective inverse temperature $\beta_{\rm eff}=\frac{\alpha\mu}{v^2}$.
    
    \item At second order in $V$, the density remains a local function of the potential, but is \textit{not} of the Boltzmann form and depends explicitly on derivatives of the potential.
    
    \item At third order and higher, the density profile is
      \textit{not a local function of the potential} $V(x)$. Namely,
      the density at point $x$ can depend, in general, on the shape of
      the potential everywhere in the system. Indeed, at cubic order
      in the potential, we have a contribution of the form $\int_0^x
      dx'\, (\partial_{x'}V)^3$. This has many striking
      consequences. For example, when an asymmetric, say sawtooth
      potential is placed in the middle of the system, it is easy to
      check using the results above that it leads
      to different densities on both of its sides, with the density
      difference depending on the exact shape of the asymmetric
      potential. This is illustrated in Fig. \ref{fig:asymmetric
        barrier} and can be thought of as a simple model for the
      bacterial ratchet experiment described in
      \shortciteNP{Galajda2007}\footnote{For a detailed explanation
        more closely related to the experiment, which also highlights
        the role of asymmetric tumbles, see~\shortciteNP{Tailleur2009EPL}.}.
      Furthermore, the same setup with periodic boundary conditions
      instead of walls leads to a non-uniform density profile with a
      current flowing in the system.
    \begin{figure}
      \begin{center}
        \includegraphics[width=.7\textwidth]{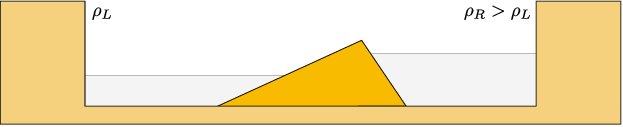}
      \end{center}
      \caption{An asymmetric barrier placed in an active system. Due to the asymmetry of the barrier, the steady-state distribution of active particles shows density differences between the two sides of the barrier. This stems from the non-local nature of the distribution, as evident from Eq.~(\ref{eq:Q}). Imposing periodic boundary conditions would generate a steady-state current through the system.}\label{fig:asymmetric barrier}
    \end{figure}

    \item The expression for $\Delta$, Eq. (\ref{eq:Delta eom}),
      implies, along with the expression for the density, that
      particles are polarized inside the potential region. This occurs even
      to leading order in $V$, when the system is effectively in
      equilibrium.
    
    \item Using the expression for the density (or methods similar to those of Sec. \ref{sec:overdamped}) it is easy to verify that, for the model defined above, $P_M$ admits an equation of state. This is not in contradiction with the non-locality mentioned above: outside the wall region, $\partial_x V=0$ and the steady state is uniform. Furthermore, the existence of an equation of state for $P_M$ is in line with the results of the previous section, as there is no analog of torques in the model. Torques can be incorporated through a position dependent flipping rate $\alpha(x)$, which indeed leads to a lack of an equation of state for $P_M$. It is important to stress that the equation of state arises in the presence of polar order inside the potential $V(x)$. In other words, there is no relation between the existence of polar order inside the potential and the lack of an equation of state. 
\end{itemize}
The above discussion again highlights that even when the system has features that are manifestly non-equilibrium, an equation of state for $P_M$ may or may not exist. Along with the results of Sec. \ref{sec:overdamped}, this suggests that active systems generically have no equation of state, but as we saw, and rather surprisingly, in some cases they do. In what follows, we will see that in systems that possess an EOS, there is a hidden ``conserved'' quantity in the steady state. While the discussion can be carried out directly for overdamped systems, it is more transparent for an underdamped model of ABPs that we introduce below.

\section{Momentum and active impulse} \label{sec:impulse}

We now return to address the question of when and why an equation of state emerges in some cases while in others it does not. As we saw in Sec.~\ref{sec:definitions}, the existence of bulk momentum conservation in the system and the existence of an equation of state are linked. To this end, in the following section, we consider a model of underdamped ABPs and study its momentum flux. Note that the following discussion is done in two dimensions.
\subsection{Underdamped active Brownian particles}

For $N$ underdamped, non-interacting active Brownian particles, the position of particle $i$, located at ${\bf r}_i$, evolves according to \shortcite{fily2017mechanical}
\begin{eqnarray}
\dot{\vec r}_i&=&\vec v_i \nonumber\ ,\\
m\dot{\vec v}_i&=&-\tilde{\gamma} \vec v_i +f_i \vec u (\theta_i)-\vec\nabla_{{\bf r}_i}
V +\sqrt{2\tilde{\gamma}^2  D_t} \boldsymbol{\eta}_i\ ,
\label{eq:dyn}
\end{eqnarray}
where $\tilde\gamma$ is the inverse mobility of the active particles,
$f_i$ their propulsive forces, and
$V$ the potential exerted by the confining walls.
$\boldsymbol{\eta}_i$'s are Gaussian white noises satisfying $\langle
{\eta}^\alpha_i(t) \rangle=0$
and $\langle {\eta}^\alpha_i(t) {\eta}^\beta_j(t') \rangle =
\delta_{ij}\delta^{\alpha\beta} \delta(t-t')$, where the angular brackets
denote an average over noise histories. ${\bf u}(\theta_i)$ is a director along the orientation, $\theta_i$, of particle $i$ which evolves according to the overdamped dynamics
\begin{equation}\label{eq:theta}
  \dot{\theta}_i= \Gamma_i({\bf r}_i,\theta_i)  + \sqrt{2D_r} \zeta_i \;.
\end{equation}
Here $\Gamma_i({\bf r}_i,\theta_i)$ is the torque (again, we silently absorb the rotational mobility in the torque) exerted on the particle by
the wall, and $\zeta_i$ is a Gaussian white noise with $\langle
\zeta_i(t) \rangle=0$ and $\langle
\zeta_i(t)\zeta_j(t')\rangle = \delta_{ij}\delta(t-t')$. Note that when $f_i=0$, the dynamics satisfy a fluctuation-dissipation relation and the system is in equilibrium.

Clearly, since, for example, every active force $f_i\vec{u}(\theta_i)$ injects momentum into the system, there is no momentum conservation in this model. Nonetheless, as suggested above, it is useful to consider the momentum density field of the model. To this end, we consider both the density, $ \hat{\rho}(\vec r)$,  and momentum density fields, $\hat{\vec p}(\vec r)$, defined through
\begin{eqnarray}
    \hat{\rho}(\vec r)&=&\sum_i \delta(\vec r-\vec r_i)\nonumber \ ,\\
    \hat{\vec p}(\vec r)&=&\sum_i  m \vec v_i \delta (\vec r-\vec r_i) \;,
    \label{eq:momentum}
\end{eqnarray}
and whose averages with respect to noise realizations and initial conditions are denoted by $\rho(\vec r)=\langle \hat \rho(\vec r)\rangle$ and $\vec p(\vec r)=\langle \hat {\vec p}(\vec r) \rangle$. 

To proceed, we consider the dynamics of the density and momentum fields. The density field obeys
\begin{equation}\label{eq:dynrho}
 \partial_t {\hat \rho}(\vec r) = \sum_i  \dot {\vec r}_i \cdot \grad_{\vec
r_i}\delta(\vec r-\vec r_i) = -\frac{1}{m}\grad \cdot \hat {\vec p}(\vec r) \;.
\end{equation}
Here and in what follows, the subscript $\vec r_i$ indicates that the gradient acts on the coordinates of the particle $i$. In the absence of a subscript, it acts on $\vec r$. The dynamics of the momentum-density field can be obtained by differentiating Eq. (\ref{eq:momentum}) and using the equations of motion~\eqref{eq:dyn}, which leads to
\begin{eqnarray}
\label{mom_dynamics}
\partial_t \vec{\hat p} &=& \sum_i\left( -\tilde{\gamma} {\vec v}_i
  -\nabla_{\vec r_i} V +f_i \vec u(\theta_i) +\sqrt{2 \tilde{\gamma}^2 D_t}
  \vec{\eta}_i\right)\delta(\vec r-\vec r_i) \nonumber \\
 &&+\sum_i  m \vec {v_i} (\vec{v_i} \cdot
\grad_{\vec r_i}) \delta(\vec r-\vec r_i)  \nonumber \\
&=& -\gamma \vec{\hat p} -  \hat \rho
\nabla V+\sum_i  f_i \vec
u(\theta_i) \delta(\vec r-\vec r_i) +\sqrt{2 \tilde{\gamma}^2 D_t \hat \rho} \,\boldsymbol{\Lambda}      -
\grad\cdot [  {\mathcal J} ] \;,
\end{eqnarray}
where $\gamma\equiv \tilde{\gamma}/m$. In the second line we define the tensor ${\mathcal J}$ through
\begin{equation}
  {\mathcal J}(\vec{r}) \equiv
  \sum_i m{\vec v_i}{\vec v_i} \delta(\vec r-\vec r_i)
\end{equation}
with ${\bf v}_i {\bf v}_i$ implying a tensor product. The Gaussian white noise is given by
$\displaystyle\sqrt{2 \tilde{\gamma}^2 D_t \hat \rho} \,\boldsymbol\Lambda\equiv$ $\sum_i$
$\displaystyle\sqrt{2 \tilde{\gamma}^2 D_t} \boldsymbol{\eta}_i
\delta({\bf r}-{\bf r_i})$
and can be verified to obey $\langle \Lambda_\alpha({\bf
  r},t) \Lambda_\beta({\bf r'},t')\rangle = \delta_{\alpha \beta}\delta({\bf r}-{\bf
  r'})\delta (t-t')$. 
The different terms in Eq. (\ref{mom_dynamics}) can be interpreted as follows:
\begin{itemize}
\item[(i)] The loss of momentum through dissipation, $-\gamma \vec{\hat p}$.
\item[(ii)] The change in momentum due to forces exerted by the walls, $- \hat \rho \nabla_x
  V$.
\item[(iii)] The change in momentum due to active forces
propelling the particles, $\sum_i f_i \vec u(\theta_i) \delta(\vec
r-\vec r_i)$.
\item[(iv)] Fluctuations, $\sqrt{2 D_t \hat \rho}\, {\bf \Lambda}$.
\item[(v)] Advection of momentum through the motion of particles arriving and
  departing from $\vec r$, $ - \grad\cdot [ {\mathcal  J} ] $.
\end{itemize}
Note that the $(\alpha,\beta)$ component of the ${\mathcal J}$ tensor,
${\mathcal J^{\alpha\beta}}=\sum_i m v_i^\alpha
v_i^\beta\delta(\vec{r}-\vec r_i)$, is the momentum flux along $\hat
\alpha$ of momentum along $\hat \beta$. Also note that only the last
term, $\nabla\cdot[\mathcal J]$, has a momentum-conserving form.

\subsection{Momentum sources and sinks}
As before, we study a system with confining walls parallel to the
$\hat{y}$ direction and with periodic boundary conditions along that
direction. In addition, for simplicity, we take the system's length
along the $\hat{y}$ direction to be equal to one.  Since this
confining potential does not allow for currents in the
system\footnote{Note that there is no mechanism in the model that can
  possibly lead to a current along the $\hat y$ direction.}, the
expectation value of the momentum fields is zero in the steady state,
${\bf p}(\vec r)=0$. Integrating Eq. (\ref{mom_dynamics}) along the
$y$ coordinate and averaging over the steady-state distribution
(denoted by angular brackets), we obtain
\begin{equation}\label{eq:localpressure}
  0 = -\rho(x) \partial_x V(x) + \bigg\langle\sum_i  f_i \cos \theta_i
  \delta(x-x_i)  \bigg\rangle - \partial_x[ \langle {\mathcal J}^{xx}(x) \rangle ] \;,
\end{equation}
where we defined $\rho(x)=\int_0^1 dy \rho (x,y)$. This expression has a simple interpretation: The momentum
flux through the system, encoded in ${\mathcal J}^{xx}$, is modulated in space by both the force exerted by the wall and the active forces.

From the expression above we can easily obtain an expression for the mechanical pressure exerted on the wall. Integrating over $x$, we find
\begin{equation}\label{eq:globalpressure}
 P_M = \int_0^\infty ds\, \rho \nabla_s V =  \langle {\mathcal J}^{xx}(0) \rangle 
+  \int_0^{\infty}ds\, \bigg\langle\sum_i  f_i \cos \theta_i \delta(s-x_i) \bigg\rangle\;,
\end{equation}
where we set $x=0$ to be a point deep in the bulk of the system. The equation implies that the total decrease in the momentum flux ${\mathcal J}^{xx}$ from its bulk value $\langle{\mathcal J}^{xx}(0)\rangle$ to zero is a result of the total force exerted by the wall and the total active force exerted in the $x>0$ region. That is, the pressure is given by the overall momentum flux entering the region $x>0$ and the total active force exerted in this region.

At a first glance, it seems that the last term of Eq. (\ref{eq:globalpressure}) suggests that there is no EOS for this model. However, in Sec. \ref{sec:Pressure of ABPs} we saw that this model, in its overdamped version, does have an equation of state in the absence of wall torques. As the overdamped limit can be taken at this point as well, this system must have an EOS when $\Gamma({\bf r},\theta)=0$. Thus, in this case, the second term of Eq. (\ref{eq:globalpressure}) must be expressible as a local bulk quantity.

To re-express this term, we note that
\begin{eqnarray}\label{eq:localrateactiveforce}
  \partial_t \bigg\langle \sum_i f_i \cos\theta_i \delta(x-x_i) \bigg\rangle &=&  -
\bigg\langle \sum_i f_i \Gamma_i \sin\theta_i \delta(x-x_i) \bigg\rangle - D_r \bigg\langle
\sum_i f_i \cos\theta_i \delta(x-x_i) \bigg\rangle\nonumber\\
  && -
\partial_x\bigg[ \bigg\langle \sum_i v_i^x f_i \cos\theta_i \delta(x-x_i) \bigg\rangle\bigg]
\end{eqnarray}
where we used Eq.~\eqref{eq:theta} and the It\=o rule for differentiation. In the steady state, this gives
\begin{equation}\label{eq:localmeanactiveforce}
  \bigg\langle \sum_i f_i \cos\theta_i \delta(x-x_i) \bigg\rangle =  - \bigg\langle \sum_i \frac{f_i}{D_r} \Gamma_i \sin\theta_i \delta(x-x_i) \bigg\rangle   - \partial_x\bigg[ \bigg\langle \sum_i \frac{v_i^x}{D_r} f_i \cos\theta_i \delta(x-x_i) \bigg\rangle\bigg]\;.
\end{equation}
The active force is decomposed into two terms -- one which depends explicitly on the torques exerted by the wall and another, divergence-like term, which does not depend explicitly on the torque. The pressure then takes the form \shortcite{fily2017mechanical}
\begin{equation}\label{eq:pressureEOStorque}
 \boxed{P_M = \langle {\mathcal J}^{xx}(0) \rangle  + \sum_i \bigg\langle \frac{v_i^x}{D_r}
f_i \cos\theta_i \delta(x_i)\bigg\rangle
-\int_0^\infty dx\, \bigg\langle \frac{f_i}{D_r} \Gamma_i \sin \theta_i \delta (x-x_i)
\bigg\rangle}
\end{equation}
where $x=0$, as above, is a point deep in the bulk of the system. The second term written above correlates the velocity in the $\hat{x}$ direction, $v_i^x$, with the $\hat{x}$ projection of the director, $u_i^x$. This term is in general non-zero even in a uniform isotropic bulk. Importantly, the first two terms depend only on the bulk properties of the system, while the third term depends on the distribution of particles inside the wall and the form of the torque. Evidently, this last term is the one responsible for the general lack of an equation of state for $P_M$. In its absence ($\Gamma=0$), $P_M$ clearly has an equation of state.

To understand this further, it is useful to rewrite Eq.~(\ref{eq:localpressure}) in the form
\begin{equation}\label{eq:forcebalance}
  \rho(x) \partial_x V(x) + \sum_i  \bigg\langle\frac{f_i}{D_r} \Gamma_i \sin\theta_i \delta(x-x_i)
  \bigg\rangle = 
  - \partial_x\Big [ \sum_i \bigg\langle \frac{v_i^x}{D_r} f_i
  \cos\theta_i \delta(x-x_i) \bigg\rangle+ \langle {\mathcal J}^{xx}(x) \rangle \Big] \;.
\end{equation}
The right-hand side of this equation is a divergence of a local
quantity, and hence has the form of a ``conserving'' piece, where the
quotations indicate that the conservation holds only in the steady
state. $\langle {\mathcal J}^{xx}(x) \rangle$ accounts for the flow of
momentum by the particles and the second term will be discussed
shortly. Note that these fluxes are balanced by both the forces
exerted by the potential and by a term related to the active forces
and torques, see the left-hand-side of
Eq.~\eqref{eq:forcebalance}. These last two terms effectively act as
sources or sinks of momentum.

\begin{figure}[ht]
\includegraphics[width=\textwidth]{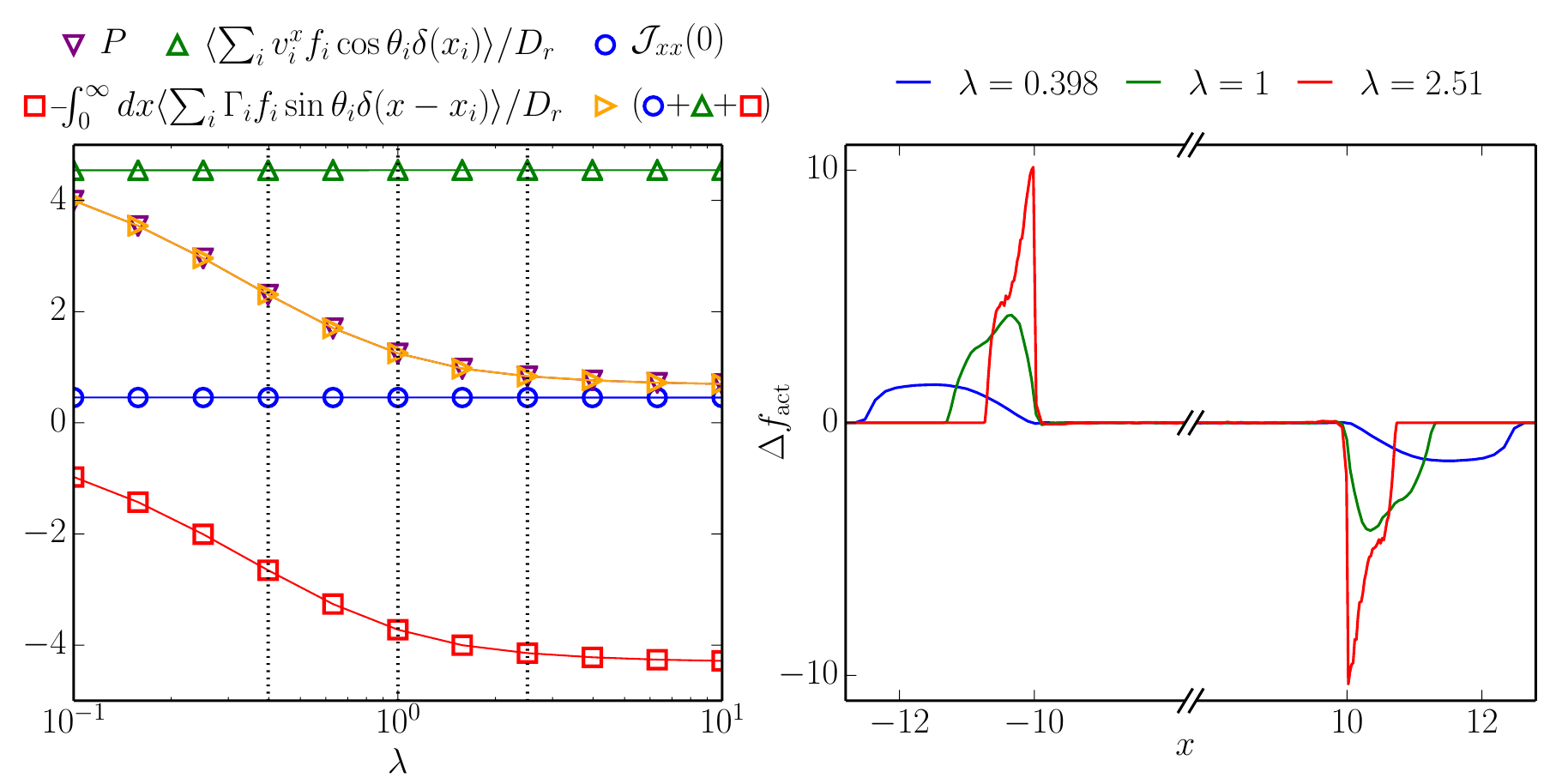}
\caption{Simulations of self-propelled ellipses for $x_w=10$, $v_0=1$,
  $D_r=0.1$, $D_t=0$, $\tilde \gamma=1$, $m=1$, $\kappa=1$, $\rho_0=1$. {\bf
    Left} Plot of the various contributions to the pressure listed in
  Eq.~(\ref{eq:pressureEOStorque}) as the wall stiffness varies. Only
  the torque-dependent term depends on the wall stiffness, leading to
  a breakdown of the EOS. $P$ is measured from its definition in
  Eq.~(\ref{eq:globalpressure}). {\bf Right} The wall-dependent
  sources and sinks, shown for three different wall stiffnesses, are
  localized in the vicinity of the walls. Figure adapted from \protect\shortciteNP{fily2017mechanical}.}\label{sourcesinks}
\end{figure}

To further illustrate these results, we turn to numerical simulations. We consider the two-dimensional system described above where the system is confined by half-harmonic
potentials on the right $V^R$ and left $V^L$ starting at $x_w$ and $-x_w$ respectively,
\begin{equation}
\label{eq:potential}
  V^R(x)=\lambda_R\frac{(x - x_w)^2}2\Theta(x-x_w)\qquad\mbox{and}\qquad   V^L(x)=\lambda_L\frac{(x + x_w)^2}2 \Theta(-x-x_w) \;,
\end{equation}
where $\Theta(x)$ denotes the Heaviside function. We treat the particles as point-like ellipses experiencing a torque of the form 
\begin{equation}
  \Gamma^R=\lambda_R\,\kappa \Theta(x-x_w) \sin 2\theta\qquad\mbox{and}\qquad  \Gamma^L=\lambda_L\,\kappa \Theta(-x-x_w) \sin 2\theta
\end{equation}
where $\kappa=\mu_r(a^2-b^2)$ is a measure of the anisotropy of the particles and $\mu_r$ is a rotational mobility \shortcite{Solon2015NatPhys}. 
The three contributions to the pressure $P_M$ defined in
Eq.~(\ref{eq:pressureEOStorque}) are shown in the left panel of
Fig.~\ref{sourcesinks} for walls with
$\lambda_R=\lambda_L=\lambda$. The figure shows that only the
torque-dependent contributions depend on the wall stiffness
$\lambda$. The corresponding wall-dependent sources and sinks $\Delta f_{\rm
  act}(x)=-\sum_i\langle\frac{f_i}{D_r} \Gamma_i \sin\theta_i \delta(x-x_i)\rangle$ are shown in the right panel of Fig.~\ref{sourcesinks}
  for three stiffness values. Clearly, the pressure varies due to the physics in the vicinity of the walls. The torques applied by the walls induce net momentum sources or sinks, which cause the breakdown of the EOS.

From the expression in Eq. (\ref{eq:pressureEOStorque}), we see that part of the active force has now been subsumed into a ``momentum-conserving''-like term: the second term of the right-hand side is indeed evaluated in the bulk of the system, much like the momentum flux ${\mathcal J}^{xx}$. As the active particles keep pumping momentum into the system, it is not clear what is the origin of this fact. In particular, in analogy with ${\mathcal J}^{xx}$ describing the flow of momentum, it is natural to ask what are the quantities whose flow is described by the second term of Eq. (\ref{eq:pressureEOStorque}). 

\subsection{The active impulse}
To answer this question, we consider the torque-free version of the
model above, where an equation of state for $P_M$ exists. All the
statements to follow generalize to cases in which the dynamics of the
active force $f_i {\bf u}(\theta_i)$ is decoupled from other degrees
of freedom~\shortcite{fily2017mechanical}. As noted above, the
contribution of the activity to the pressure stems from the momentum
transfer to the particles through the active forces. Hence, it is
natural to look at how much momentum the active particle will receive,
on average, from its active force in the future. This is directly
quantified by the \textit{active impulse}:
\begin{equation}\label{eq:activeimpulse}
  \Delta {\bf p}_i^{\rm a}(t)\equiv \int_t^\infty \overline{f_i {\bf
      u[\theta_i(s)]}} ds = \frac{f_i}{D_r} {\bf u}[\theta_i(t)]
\end{equation}
where the over-bar denotes an average with respect to histories of the
system in the time interval $[t,+\infty)$ for a fixed value of
  $\theta_i(t)$.  In~(\ref{eq:activeimpulse}), the active impulse
  simply depends on the initial angle $\theta_i(t)$ because the
  dynamics of the active force $f_i {\bf u}(\theta_i)$ is independent
  of all other degrees of freedom, and is randomized by the rotational
  noise after a time $1/D_r$, {\it even where the potential $V({\bf
      r})$ is non-zero}. Note that, by construction, the active
  impulse obeys
\begin{equation}\label{eq:activeimpulsedyn}
    \partial_t \Delta {\bf p}_i^{\rm a}(t)= - f_i {\bf u}[\theta_i(t)] 
     \;.
\end{equation}

Turning from the single-particle description to a many-body description, we define an active impulse field, $\langle
\Delta {\bf p^{\rm a}}(x)\rangle = \langle \sum_i\Delta {\bf p}_i^a
\delta({\bf r}-{\bf r}_i) \rangle$.
Using Eq. (\ref{eq:activeimpulsedyn}) in the steady state, this field satisfies
\begin{equation}\label{eq:dynFID}
0 = \partial_t \langle \Delta {\bf p^{\rm a}}(x) \rangle= - \bigg\langle \sum_i f_i {\bf
u}[\theta_i(\vec{s})] \delta({\bf r}-{\bf r}_i) \bigg\rangle- \nabla \cdot \bigg\langle \sum_i
{\bf v}_i \Delta{\bf  p^{\rm a}}_i \delta({\bf r}-{\bf r}_i) \bigg\rangle\ ,
\end{equation}
where the divergence $\nabla$ is contracted with the velocities ${\bf v}_i$. Note that the first term on the right-hand side of this equation represents the decay of $\Delta {\bf p}^a$ due to the production of a non-zero mean active force; in the steady state, it has to be balanced by the second term which is the influx of $\Delta {\bf p}^a$. Re-expressing the active impulse using Eq. (\ref{eq:activeimpulse}), we find 
\begin{eqnarray}
\bigg\langle \sum_i f_i {\bf u}[\theta_i(\vec{s})] \delta({\bf r}-{\bf r}_i) \bigg\rangle &=& -
\nabla \cdot \bigg\langle \sum_i
{\bf v}_i \Delta{\bf p}^{\rm a}_i \delta({\bf r}-{\bf r}_i) \bigg\rangle\nonumber\\&=& -
\nabla \cdot \bigg\langle \sum_i
{\bf v}_i \frac{f_i}{D_r}{\bf u}(\theta_i) \delta({\bf r}-{\bf r}_i) \bigg\rangle \;.\label{activeforcediv}
\end{eqnarray}

\begin{figure}
    \begin{center}
        \includegraphics[width=.99\textwidth]{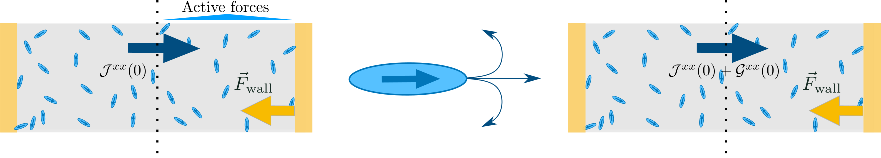}
      \end{center}
    \caption{{\bf Left:} Particles enter the right half of the system carrying their momentum $m v_i$. The corresponding flux of momentum, ${\mathcal J}^{xx}$, will fall to zero outside the container, since there are no particles there. This decrease of the momentum flux is due to the conjugated effect of the wall force $F_w$ and the active forces (see Eq.~\eqref{eq:globalpressure}). {\bf Center:} An active particle randomizes its force after a typical time $D_r^{-1}$. Particles deep in the right side of the system thus do not receive any momentum from their active forces on average. The total momentum the active particles exchange with the substrate in the right side of the system is therefore given by the active impulse they bring with them as they cross the boundary at $x=0$. See Eq.~\eqref{eq:forcebalancewithG}.  {\bf Right:} The force exerted by the wall is thus balanced by the sum of the incoming fluxes of momentum and active impulse, which leads to an EOS for the pressure in the absence of torques~\eqref{eq:pressureEOS}.}
    \label{fig:AI}
\end{figure}

The above relation shows that, despite the fact that active forces inject momentum into the system, in the steady state, their average contribution takes the form of a divergence of a local quantity. We refer to this quantity as the active-impulse flux tensor, ${\mathcal G}({\bf r})$, whose definition is given by
\begin{equation}\label{eq:activeimpulsetensor} {\mathcal G}({\bf r})\equiv
  \sum_i {\bf v}_i \frac{f_i}{D_r}{\bf u}(\theta_i) \delta({\bf
    r}-{\bf r}_i)\;.
\end{equation}
Equation~\eqref{activeforcediv} can be understood as follows: since
$\theta_i$ undergoes rotational diffusion, the contribution of an
active particle entering a given volume decays to zero with time. The
only way for the mean local active force to sustain its non-zero value
is thus by incoming fluxes of particles entering the volume, each
carrying its own active force (see Fig.~\ref{fig:AI}). As $\vec{v}_i$
and $\vec{u}_i$ are correlated, such fluxes are non-vanishing.

Inserting the definition of the flux of the active impulse tensor
$\mathcal{G}$ into Eq.~(\ref{eq:forcebalance}), and setting
$\Gamma=0$, we find
\begin{eqnarray}
\label{eq:forcebalancewithG}
  \rho(x) \partial_x V(x) &=& 
  - \partial_x\Big [ \sum_i \bigg\langle \frac{v_i^x}{D_r} f_i
  \cos\theta_i \delta(x-x_i) \bigg\rangle+ \langle {\mathcal J}^{xx}(x) \rangle \Big] \nonumber\\
  &=& - \partial_x\Big [\langle {\mathcal G}^{xx}(x) \rangle + \langle {\mathcal J}^{xx}(x) \rangle \Big]\; ,
\end{eqnarray}
and so the pressure takes the form
\begin{equation}\label{eq:pressureEOS}
  P_M =
\langle {\mathcal J}^{xx}(0) \rangle  + \langle {\mathcal G}^{xx}(0)\rangle.
\end{equation}
\hfill \break
With this result, we make the following comments:
\begin{itemize}
    \item Using It\=o formula, one directly shows that
\begin{eqnarray}
 \partial_t \bigg\langle\sum_i{\vec{r}}_i f_i {\bf u}(\theta_i)\delta({\vec{r}}-\vec{r}_i)\bigg\rangle&=& \bigg\langle\sum_i\vec{v}_i f_i {\bf u}(\theta_i)\delta({\vec{r}}-{\vec{r}}_i)\bigg\rangle - D_r\bigg\langle\sum_i{\vec{r}}_i f_i{\bf u}(\theta_i)\delta(\vec{r}-\vec{r}_i)\bigg\rangle \nonumber\\
  && - \nabla\cdot \bigg\langle\sum_i\dot{\vec{r}_i}\vec{r}_i f_i{\bf u}(\theta_i)\delta(\vec{r}-\vec{r}_i)\bigg\rangle\; .
\end{eqnarray}
     The steady-state average of $\mathcal G$ in a homogeneous system can thus be identified with the ``swim pressure'', introduced by Takatory and Brady \shortcite{Takatori2014} and by Yang, Manning and Marchetti \shortcite{Yang2014}:
       \begin{equation}\label{eq:swimpressure}
        \langle\mathcal{G}(\vec{r})\rangle = \bigg\langle\sum_i \delta(\vec{r}-\vec{r}_i)\vec{r}_i f_i\vec{u}(\theta_i)\bigg\rangle \;.
    \end{equation}
    Given the lack of an explicit solvent in our description, we prefer the term ``active pressure'' for this contribution. This validates the virial-based approaches in this context~\shortcite{Winkler2015SoftMatter,falasco2016mesoscopic}.
    The above discussion shows that the swim-pressure is the flux of active impulse. Note, however, that it should now be clear that this expression does not, in general, provide the expression for $P_M$, when $\Gamma({\bf r},\theta)\neq 0$ for example. Furhtermore, Eq.~\eqref{eq:swimpressure} is only valid in the steady state of homogeneous systems.
    
    \item Recall that all of the discussion above is done in the steady state. Averaging over Eq. (\ref{mom_dynamics}) in the steady state and using the results presented above gives
    \begin{equation}
        0 = \partial_t \langle\vec{p}\rangle = -\gamma\langle\vec{p}\rangle - \partial_t \langle\Delta\vec{p}^a\rangle - \langle\rho\rangle \nabla V - \nabla\cdot[\langle\mathcal{G} + \mathcal{J}\rangle] \;,
    \end{equation}
    or alternatively
    \begin{equation}
        0 = \partial_t \langle\vec{p} + \Delta\vec{p}^a\rangle = -\gamma\langle\vec{p}\rangle - \langle\rho\rangle \nabla V - \nabla\cdot[\langle\mathcal{G} + \mathcal{J}\rangle] \;.
    \end{equation}
    We therefore find that in a flux-free steady state, and outside
    the wall, $(\langle\vec{p}\rangle+\langle\Delta\vec{p}^a\rangle)$
    is conserved. This is the hidden ``conserved'' quantity which is
    responsible for the appearance of an equation of state for $P_M$.
    Note that this ``conservation'' is not guaranteed for other cases
    -- outside of the steady state and even for curved walls, which
    are known to generate currents, with $\langle\vec{p}\rangle \neq 0$.
  
      \item Generalizing this treatment to include interactions does not modify the important points of the discussion above. We refer the reader to \shortciteNP{fily2017mechanical} for details.
\end{itemize}

Finally, note that even in cases in which an EOS exists ($\Gamma=0$), the ``conservation'' of $(\langle\vec{p}\rangle+\langle\Delta\vec{p}^a\rangle)$ rests on the fact that the expectation value of the momentum field vanishes -- $\langle\vec{p}\rangle=0$. But, as seen in Sec. \ref{sec:steady-state}, the one-dimensional configuration illustrated in Fig. \ref{fig:asymmetric barrier} generates currents in the system, with $\langle\vec{p}\rangle\neq0$. It is rather generic for currents to exist in an active system and it is thus natural to ask how currents modify the picture described above.

 \section{Objects immersed in an active bath: Currents and forces} \label{sec:currents}
\begin{figure}
  \begin{center}
    \includegraphics[width=.6\textwidth]{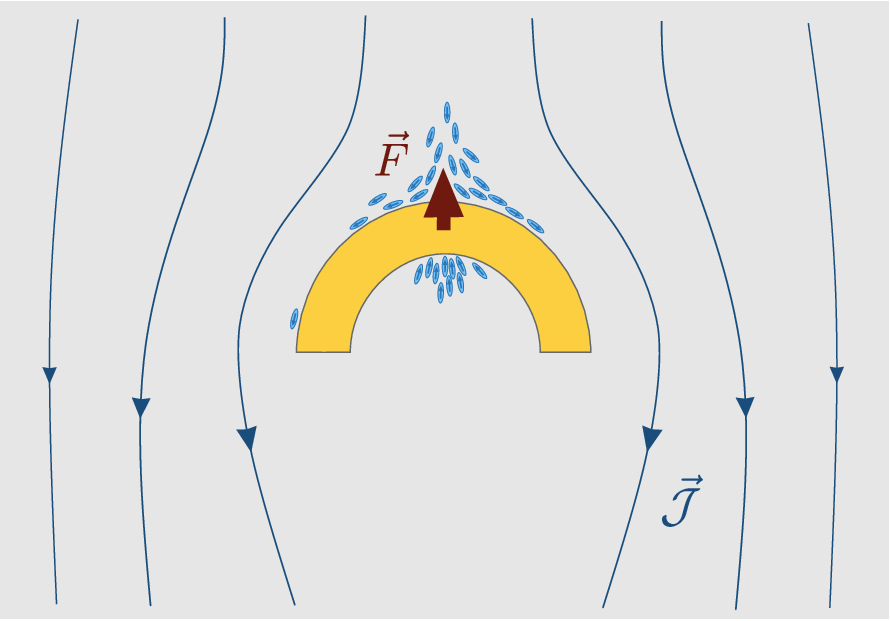}
  \end{center}
  \caption{An asymmetric object, here a semi-circle, is placed in an active medium. As discussed above, such an object induces a current in the system, which, in turn, exerts a force back on the object. This force is opposite in direction to the total current, as active particles accumulate on its concave side of the semi-circle and glide off its convex side.}\label{fig:dipole flow}
\end{figure}

Another situation in which the statistics of active forces play an
important role is when passive objects are immersed in an active
bath~\shortcite{Kikuchi24112009,ActivatingMembranes,Ratchet_reich,yan2015force,PolymerLooping,PolymerCrowdedDynamics,PolymerCollapse,Kaiser2014PRL,PolymerSwelling,PolymerBrush,smallenburg2015swim,Bechinger2016RMP,Nikolai2016PRL,junot2017active,razin2017generalized,baek2018generic}. As hinted by the final discussion of
the previous section, this can be particularly interesting when
currents are present in the system, a case on which we now focus. All
the results presented in this section have been derived
in~\shortciteNP{Nikolai2016PRL} and~\shortciteNP{baek2018generic}. As argued above,
currents naturally arise when a non-symmetric potential -- a passive
asymmetric object -- is placed inside an active fluid. If the interaction between
the active particles and the object lead to a particle current
resulting from a non-zero net force, Newton's third law implies that
the active particles will in turn exert a force back on the
object. This is illustrated in Fig.~\ref{fig:dipole flow} where, as an
example, a semi-circular object is placed in a two-dimensional fluid
of active particles. Due to the persistence of the active particles,
we expect them to accumulate on the concave side (inside the
semi-circle), and to glide along the convex side. This in turn
generates a current in the system whose direction is opposite to the
force exerted on the semi-circle. As we now show, this intuitive
connection between currents and forces can be shown to manifest itself
for a class of systems, including ABPs, through a simple relation
between the current surrounding an object and the force acting on it
\begin{equation}\label{eq:J=muF}
    \vec{\mathcal{J}} = - \mu \vec{F}^{(tot)} \;.
\end{equation}
Here $\vec{F}^{(tot)}=\int d{\bf r}\, \rho ({\bf r}) \nabla V({\bf r})$ is the total active force on the object, $\mu$ is the mobility of the active particles, ${\mathcal J}=\int d{\bf r}\, J ({\bf r})$ is the total current flowing through the system, and $J({\bf r})$ is the active particle current density. 

To derive this relation we use the overdamped model of non-interacting ABPs from Sec. \ref{sec:Pressure of ABPs} in two dimensions in the absence of  torques. Note that the following treatment can be generalized to models with pairwise interactions. In the absence of torques, the Fokker-Planck equation for this model takes the form 
\begin{eqnarray}
    \partial_t \mathcal{P}(\vec{r},\theta) = -\nabla \cdot [v{\bf u}(\theta)\mathcal{P}(\vec{r},\theta) - \mu (\nabla V) \mathcal{P}(\vec{r},\theta) - D_t\nabla \mathcal{P}(\vec{r},\theta)] +D_r \partial^2_\theta \mathcal{P}(\vec{r},\theta) \;.\;
\label{eq:FP_for_current}\end{eqnarray}
Integrating over $\theta$, we obtain
\begin{equation}
    \partial_t \rho(\vec{r}) = - \nabla\cdot[v m_{x,1}\hat{x} + v m_{y,1}\hat{y} - \mu\nabla V \hat{\rho}(\vec{r}) - D_t\nabla\hat{\rho}(\vec{r})] \equiv - \nabla \cdot\vec{J}
\end{equation}
where similarly to before, we define the moments $m_{\alpha,n}$ as
\begin{eqnarray}
  m_{x,n}(\vec{r}) &=& \int d\theta\,\cos(n\theta)\mathcal{P}(\vec{r},\theta) \nonumber\ ,\\
  m_{y,n}(\vec{r}) &=& \int d\theta\,\sin(n\theta)\mathcal{P}(\vec{r},\theta)\ ,
\end{eqnarray}
and $\rho(\vec{r})=m_{x,0}$. Integrating the current over a surface $s$ within the system containing the object, we find
\begin{eqnarray}\label{eq:current_inside_object}
  \vec{\mathcal{J}} &=& \int_s d^2r\,\vec{J} \nonumber\\
  &=& - \mu\int_s d^2r\,\rho\nabla V + \int_s d^2r\,[v m_{x,1}\hat{x} + v m_{y,1}\hat{y} - D_t\nabla\rho] \;.
\end{eqnarray}
Note that the first term on the right-hand side in
Eq. (\ref{eq:current_inside_object}) is the total force
$-\vec{F}^{(tot)}$ acting on particles located inside $s$ times the
mobility $\mu$. To simplify further, we multiply
Eq. (\ref{eq:FP_for_current}) by $\cos\theta$ and integrate over
$\theta$, to find
\begin{eqnarray}
  m_{x,1} &=& - \frac{1}{D_r}\nabla\cdot\bigg[\frac{v}{2} (\rho + m_{x,2})\hat{x} + \frac{v}{2} m_{y,2}\hat{y} - \mu m_{x,1}\nabla V - D_t\nabla m_{x,1}\bigg] \nonumber \\
  &\equiv& - \nabla\cdot \mathcal{M}_{x,1} \;,
\end{eqnarray}
for a wall parallel to the $\hat{y}$ direction. Similarly, one can obtain an equation for $m_{y,1}$ after proper permutations of indices
\begin{equation}
    m_{y,1} = - \nabla\cdot\mathcal{M}_{y,1}\; .
\end{equation}
Evidently, both $m_{x,1}$ and $m_{y,1}$ can be written as the divergence of a local quantity.
Using Stokes' theorem, with the surface $s$ being large enough so that its boundaries are far from the object in the homogeneous bulk, we find
\begin{equation}
  \int_s d^2r\, m_{x,1}\hat{x} = - \oint_{\partial s}dl\,\hat{n}\cdot \mathcal{M}_{x,1}=0\ ,
\end{equation}
where $\hat{n}$ is a normal to the surface $s$. The vanishing of the last integral simply stems from the fact that, in the bulk, $\mathcal{M}_{x,1}=\text{const}$. Similarly, the surface integrals of $m_{y,1}$ and $\nabla\rho$ vanish. Using all this in Eq.~(\ref{eq:current_inside_object}) proves Eq. (\ref{eq:J=muF}).

\hfill \break
Several comments are in place:
\begin{itemize}
    \item The derivation can be generalized to models with pairwise interactions between active particles, yielding the same result~\shortcite{Nikolai2016PRL}.
    \item One can show that to leading order in the far-field limit, the density is given by~\shortcite{baek2018generic}
    \begin{equation}
        \rho(\vec{r}) = \rho_{\text{bulk}} - \frac{\mu}{2\pi D_{\text{eff}}}\frac{\vec{r}\cdot\vec{F}^{(tot)}}{r^2} + \mathcal{O}(r^{-2})
    \end{equation}
    with the force exerted on the particles, $\vec{F}^{(tot)}$, given again by
    \begin{equation}\label{eq:F^tot}
      \vec{F}^{(tot)}=\int d^2r'\,\rho\nabla V \ .
    \end{equation}
    The density profile induces a diffusive current $\vec{J}$ whose expression reads
    \begin{equation}
        \vec{J}(\vec{r}) \simeq - \frac{\mu}{2\pi}\bigg[\frac{\vec{F}^{(tot)}}{r^2} - \frac{2(\vec{r}\cdot\vec{F}^{(tot)})\vec{r}}{r^4}\bigg] \; .
    \end{equation}
    This has the functional form of a dipolar field. Again we see that the density profile is a non-local function of the potential.
    \item If the object is mobile, the force exerted by the active particles will make it move. This was observed experimentally, with a passive wedge emersed in a {\it B. subtilis} suspension~\shortcite{Kaiser2014PRL}.
    \item The discussion above shows that asymmetric objects induce currents, and one could thus wonder about the impact of structured walls. There are two different cases to consider here.
    \begin{itemize}
        \item \textit{Walls with ``up-down'' symmetry:} Numerical
          measurements of the density and currents are shown in
          Fig. \ref{fig:AS walls density} for a wall of periodicity
          $L_p$ that is up-down symmetric. In such cases, one can
          define an average pressure in the $\hat{x}$ direction
        \begin{equation}\label{eq:P_x AS walls}
            \langle P_x\rangle = \frac{1}{L_p}\int_0^{L_p} dy\,P_x(y)
        \end{equation}
        using the following definition
        \begin{equation}
            P_x(y) = \int_0^\infty dx\,\rho\partial_x V\ .
        \end{equation}
        One then finds that despite the fact that $P_x(y)$ depends both on
        the details of the potential and the exact value of $y$,
        $\langle P_x\rangle$ obeys the same EOS as it would obey if the walls were flat.
        As before, the $\hat{y}$ direction was taken as the
        periodic direction and $x=0$ is a point in the bulk.  As one
        may expect, the pressure is not uniform along the wall and
        depends on the shape of the wall potential (see
        Fig. \ref{fig:AS walls P}). This result holds for
        non-interacting and pairwise interacting models.
        \begin{figure}
          \begin{center}
            \includegraphics[width=.95\textwidth]{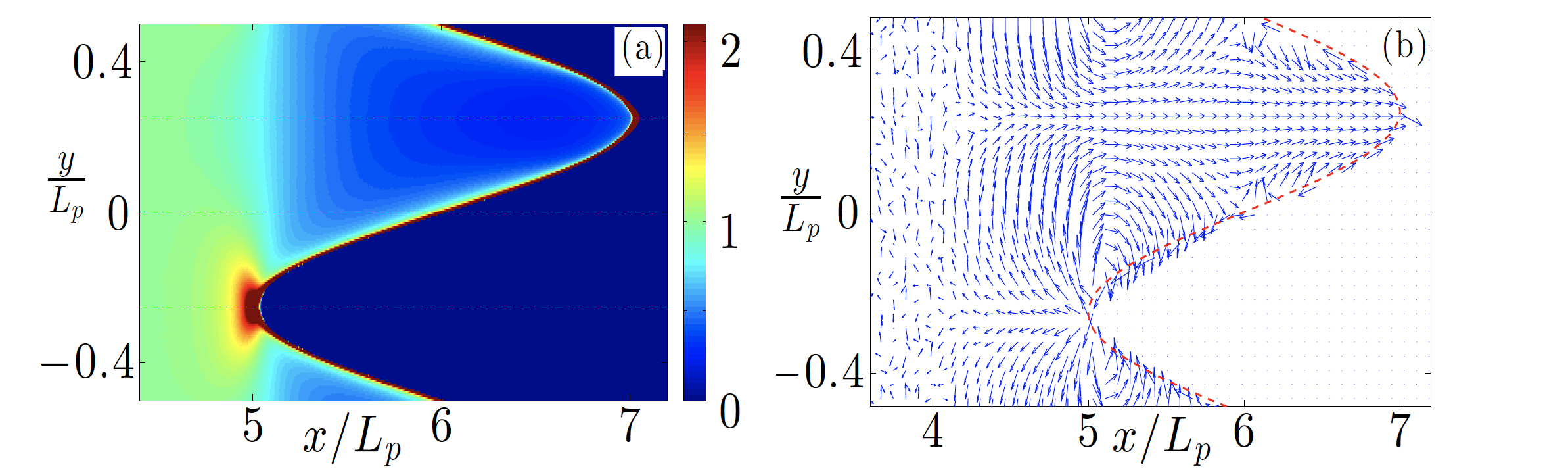}
          \end{center}
          \caption{Density (a) and current (b) of non-interacting ABPs
            near the right edge of the system with a sinusoidal hard
            wall potential of spatial periodicity $L_p=1$, spatial
            amplitude of $A=1$. The red dashed curve corresponds to
            the position of the wall. In the left panel, the color
            encodes the density, which falls quickly to zero in the
            wall region. Figure adapted from
            \protect\shortciteNP{Nikolai2016PRL}, where more details
            about the simulations are provided.}\label{fig:AS walls
            density}
        \end{figure}

        \begin{figure}
          \begin{center}
           \includegraphics[width=.5\textwidth]{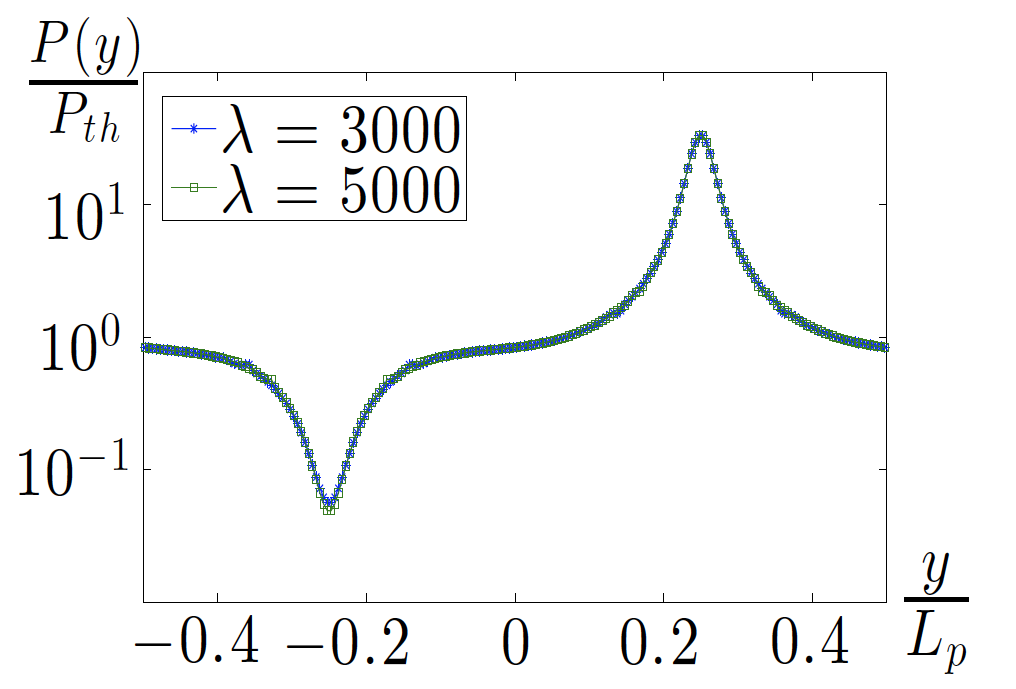}
          \end{center}
          \caption{Pressure normal to the wall, normalized by $P_{th}=\rho_0\big[\frac{v^2}{2\mu D_r}+\frac{D_t}{\mu}\big]$, the pressure of spherical ABPs near a flat wall (\ref{eq:P_M}). $\lambda$ is the  stiffness of the wall. The pressure is displayed as a function of $y$ along the wall, in the hard wall regime. Figure adapted from \protect\shortciteNP{Nikolai2016PRL}.}\label{fig:AS walls P}
        \end{figure}

        \item \textit{Walls with no ``up-down'' symmetry: } If the up-down symmetry is broken, as illustrated in Fig. \ref{fig:AS no UD walls}, the discussions above suggest that a current will be generated along the walls. This phenomenon is indeed observed numerically in Fig. \ref{fig:AS no UD walls currents}. These currents, via the discussion above, are associated with shearing forces acting on the walls. Indeed, if the wall was allowed to move, it would do so. This can be thought of as a toy model for the observed rotations of a ratchet wheel in a bacterial bath~\shortcite{DiLeonardo2010PNAS,Sokolov2010PNAS}. We note that in this case, even though $\langle P_x\rangle$ as defined in Eq. (\ref{eq:P_x AS walls}) still obeys an EOS, the force in the $\hat{y}$ direction does not and is non-zero. Thus, there is no EOS describing the stresses in this system.
        
        \begin{figure}
          \begin{center}
            \includegraphics[width=.35\textwidth]{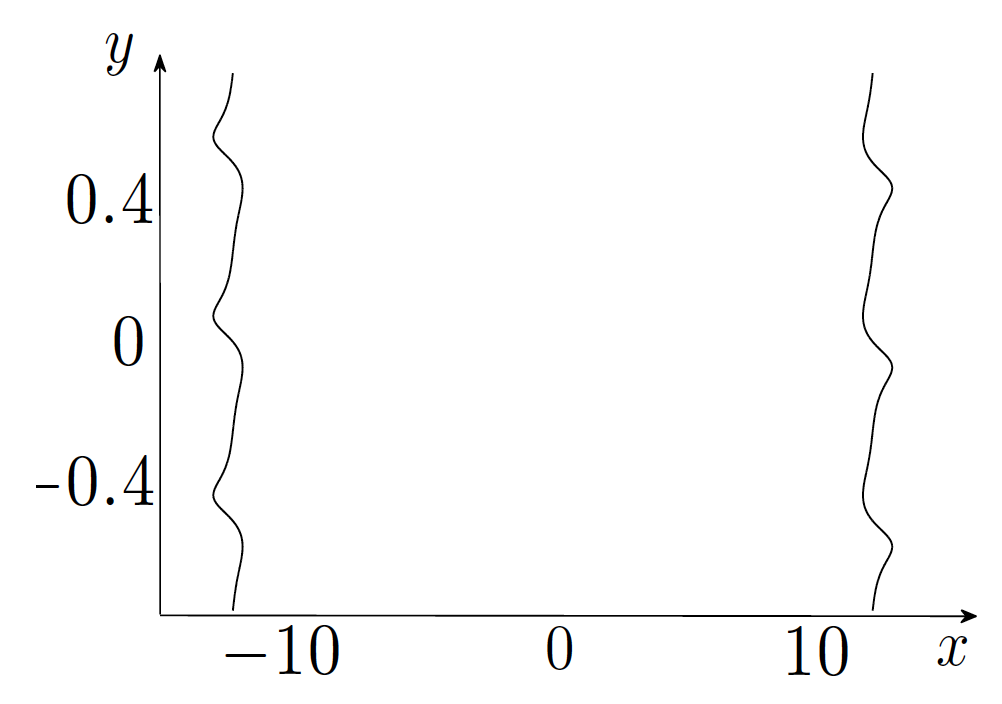}\hspace{.5cm}
            \includegraphics[width=.45\textwidth]{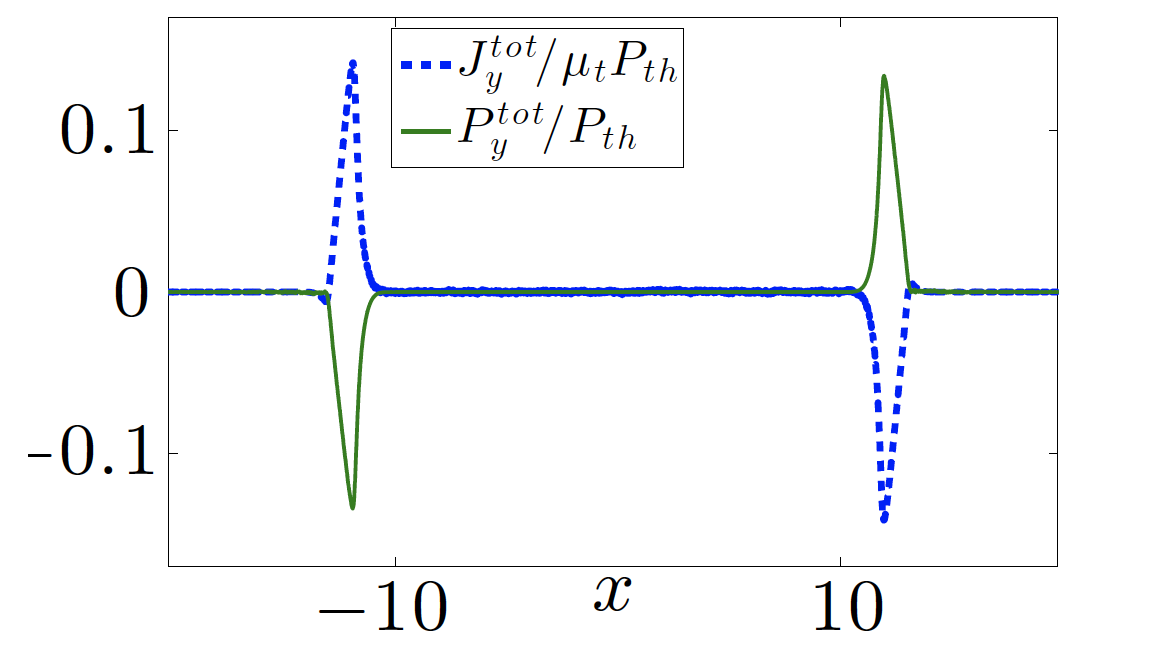}
          \end{center}
          \caption{{\bf Left:} Equipotential line of an asymmetric confining potential. As the walls are asymmetric, currents run along them. Since currents on one side will flow in the opposite direction of the current on the other wall, shear forces will be exerted on the system as a whole. {\bf Right:} Ratchet current and shear stress are functions of $x$, for such asymmetric walls. The relation between the total force and the overall current (\ref{eq:J=muF}) is verified numerically within 1\%.  Figures adapted from~\protect\shortciteNP{Nikolai2016PRL}.}\label{fig:AS no UD walls} \label{fig:AS no UD walls currents}
        \end{figure}
    \end{itemize}
\end{itemize}

\section{Acknowledgements}
In these lecture notes, we have presented works which have been done in collaboration with a number of colleagues that we thank warmly: Y. Baek, A. Baskaran, M. E. Cates, Y. Fily, M. Kardar, N. Nikola, A. P. Solon, J. Stenhammar, A. Turner, R. Voituriez, R. Wittkowski, X. Xu. YK acknowledge support from I-CORE Program of the Planning and Budgeting Committee of the Israel Science Foundation and an Israel Science Foundation grant. JT is funded by ANR Bactterns. JT \& YK acknowledge support from the MLB center for theoretical physics and a joint CNRS-MOST grant.

\bibliographystyle{OUPnamed}
\bibliography{main}

\end{document}